\documentclass{article}

\usepackage{microtype}
\usepackage{graphicx}
\usepackage{subcaption}
\usepackage{booktabs}
\usepackage{hyperref}

\usepackage[accepted]{icml2026}

\usepackage{amsmath}
\usepackage{amssymb}
\usepackage{mathtools}
\usepackage{cleveref}
\usepackage{multirow}
\usepackage{xcolor}

\definecolor{mydarkblue}{rgb}{0,0.08,0.45}
\hypersetup{ %
pdftitle={},
pdfsubject={ICML 2026 Workshop on Machine Learning for Audio},
pdfkeywords={},
pdfborder=0 0 0,
pdfpagemode=UseNone,
colorlinks=true,
linkcolor=mydarkblue,
citecolor=mydarkblue,
filecolor=mydarkblue,
urlcolor=mydarkblue,
}

\icmltitlerunning{Evaluating Pretrained Music Embeddings for Cross-Performance Jazz Standard Recognition}

\begin{document}

\twocolumn[
  \icmltitle{Evaluating Pretrained Music Embeddings for \\Cross-Performance Jazz Standard Recognition}

  \begin{icmlauthorlist}
    \icmlauthor{Cagri Eser}{me}
  \end{icmlauthorlist}
  \icmlaffiliation{me}{Department of Computer Engineering, Middle East Technical University, Ankara, Turkey}
  \icmlcorrespondingauthor{Cagri Eser}{cagri.eser@ceng.metu.edu.tr}

  \icmlkeywords{music information retrieval, jazz standard recognition, audio embeddings}

  \vskip 0.3in
]

\printAffiliationsAndNotice{}

\begin{abstract}
Recognizing jazz standards from audio is a challenging form of tune-level music retrieval: different performances of the same standard may vary in tempo, key, arrangement, instrumentation, improvisational content, and even whether the head melody is present. We study this problem using a curated subset of the Jazz Trio Database designed for cross-performance standard recognition. We compare a from-scratch trained Harmonic CNN baseline against frozen pretrained music representations from recent music understanding foundation models, using both supervised probing and nearest-neighbor retrieval. 
Our results suggest that from-scratch spectrogram models overfit strongly to training performances, while pretrained embeddings provide better top-$k$ results but are sensitive to performer identity, which can be partially reduced with a lightweight contrastive projection. Our findings motivate jazz standard recognition as a useful stress test for music representation models and as a step toward retrieval-based standard identification. Project page: \url{https://github.com/cagries/tipofmyear}.
\end{abstract}

\section{Introduction}

Audio recognition systems such as Shazam \cite{wang2003industrial} are highly effective for identifying exact recordings, but recognizing the underlying tune across different performances is a different problem. This distinction is especially important in jazz: a standard such as \emph{Autumn Leaves} can appear in many keys, different tempo, different arrangements, and in many improvisational contexts. 
In this paper, we study whether modern audio representations support this form of cross-performance tune recognition. The task is difficult for several reasons. First, jazz performances often devote long stretches to improvisation, where cropped local windows may not contain the head melody. Second, different standards can share common harmonic progressions or similar melodic fragments. Third, recordings by the same group can be acoustically and stylistically similar across performances of different standards, which causes problems for nearest-neighbor retrieval methods.

Our work makes three contributions. \textbf{First}, we construct a curated filtered benchmark subset from the Jazz Trio Database \cite{cheston2024jtd} for standard recognition across performances. \textbf{Second}, we compare spectrogram-based training, supervised probing of frozen embeddings, and embedding-based retrieval under the same evaluation protocol. \textbf{Third}, we analyze retrieval-based classification and demonstrate that nearest neighbors often retrieve performer identity rather than tune identity, and explore a lightweight supervised contrastive projection to reduce performer-biased retrieval.

\section{Related Work}

Fingerprinting systems for audio such as Shazam \cite{wang2003industrial} are very effective for exact recording identification from short and noisy excerpts, however they are designed to match a given signal to a recording already present in the database rather than recognizing a new performance of the same underlying composition.
With this in mind, jazz standard recognition is therefore closer to cover or version identification \cite{serra2011community, xun2023discover}, which is concerned with retrieval of different renditions of the same work using features that are invariant to changes in key, tempo, timbre, and arrangement. Recent version-identification systems have increasingly adopted embedding-based retrieval formulations to improve scalability while preserving work-level invariance \cite{xun2023discover}.

Recent developments in music foundation models further motivate our approach. Earlier work showed that pretrained audio representations can transfer well across downstream MIR tasks with shallow heads or no fine-tuning, including codified-audio language-model features from Jukebox-based MIR \cite{castellon2021jukemir} and broader comparative studies of supervised and unsupervised music embeddings \cite{mccallum2022audiorepr}.
MERT \cite{li2024mert} scales masked self-supervised pretraining with acoustic and musical teachers and reports strong transfer across various music-understanding tasks,
while MuQ uses Mel quantization targets and shows broad downstream gains with improvements with larger-scale pretraining. 
A closely related reference in approach is \citet{papaioannou2025universal}, who evaluate multiple audio foundation models through probing and lightweight fine-tuning across culturally diverse tagging datasets. Our setting is complementary because it probes jazz tune identity across performances rather than tag prediction across cross-cultural musical traditions.

This work complements previous approaches in the literature by using jazz standards as a deliberately hard retrieval target where unlike fingerprinting the goal is not exact recording recognition, and unlike generic cover-song benchmarks many query windows may not contain an explicit statement of the melody. Methodologically, it is closest to recent evaluation-first studies of music foundation models that compare probing and lightweight adaptation, however our target is cross-performance jazz standard identification under heavy improvisation.

\section{Dataset, Evaluation and Experiment Setup}

\begin{figure}[hbt]
\centering
\includegraphics[width=.99\linewidth]{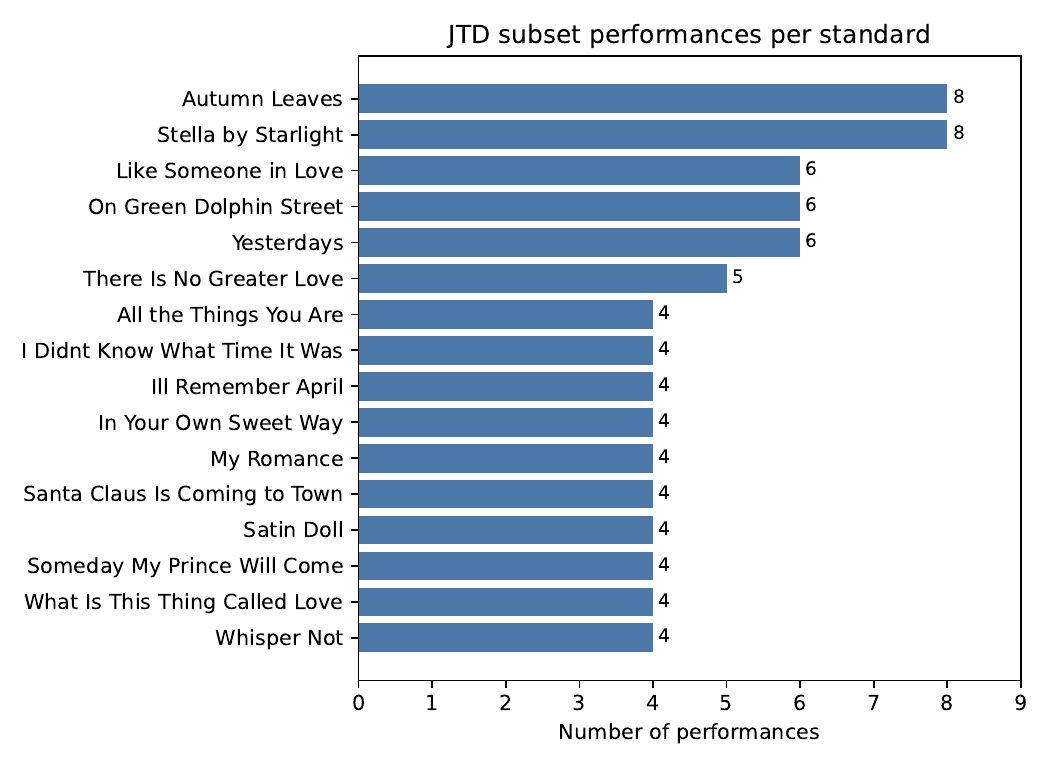}
\caption{Distribution of performances across the curated JTD subset.}
\label{fig:jtd-balanced}
\end{figure}

\paragraph{Dataset construction.}
Given our focus on jazz performances, we use the Jazz Trio Database \cite{cheston2024jtd} and curate a subset of the data intended for standard-level recognition. The JTD dataset has 1294 performances from various standards, however the distribution of the data is long-tailed: more than 85\% of standards from the dataset have fewer than 3 performances from distinct bandleader/pianist groups. From this observation, we construct a filtered dataset based on the following approach: we first canonicalize standard titles to merge spelling and naming variants. Afterwards, we collect all standards with \textbf{at least} 4 performances from distinct groups.
This threshold is chosen to support our leave-one-performance-per-standard evaluation so that in each fold one performance of every standard is held out for testing, while the remaining performances provide training and validation examples.
Finally, for each standard, we keep \textbf{at most one performance} from any given group, reducing leakage from repeated performances by the same ensemble.
The resulting training subset is not perfectly class-balanced, but it greatly reduces the long-tailed structure of the original dataset and ensures multiple performances and a balanced test setup.
Figure~\ref{fig:jtd-balanced} shows counts of various standards within the subset.
With these adjustments, the evaluation dataset contains 16 standards with 79 performances, with more metadata described in Table~\ref{tab:dataset}. 
For each selected recording, we assign a canonical standard label and segment the audio into fixed-length windows. We use the fixed standards list for all reported experiments.

\begin{table}[t]
\caption{Statistics for the curated JTD standard recognition subset.}
\label{tab:dataset}
\centering
\begin{tabular}{lr}
\toprule
Statistic & Value \\
\midrule
Standards & 16 \\
Performances & 79 \\
Performances per standard & 4.93 \\
Unique Groups & 27 \\
Window length / hop & 10 seconds / 5 seconds \\
\bottomrule
\end{tabular}
\end{table}

\paragraph{Windowing strategy.}
Each recording is converted to 24 kHz mono audio and split into 10-second windows with 5-second hop. Each window inherits the standard label of its parent performance. This creates many training examples per performance, but these windows are highly correlated within a recording. Therefore, we report performance-level metrics in addition to window-level metrics.

\begin{figure*}[t]
    \centering
    \includegraphics[width=0.99\linewidth]{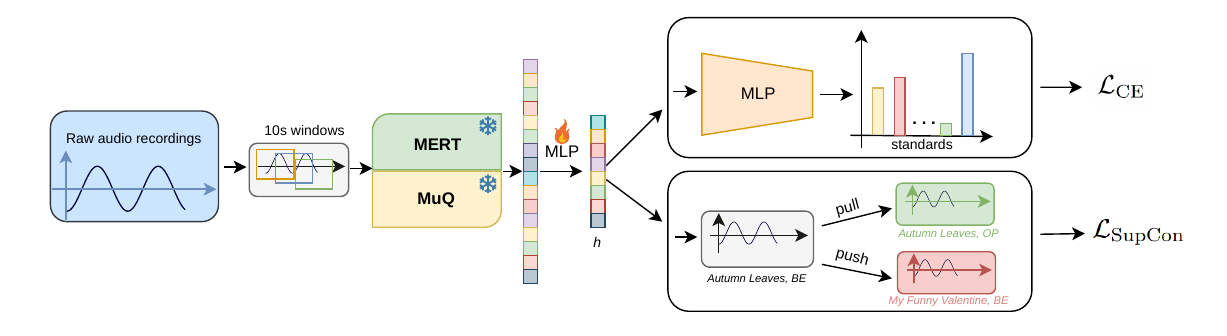}
    \caption{Pipeline for the proposed standard-aware supervised contrastive retrieval approach. Frozen MERT/MuQ embeddings are projected into a retrieval space trained to pull together windows from the same standard across different performances while reducing the same-performer retrieval bias.}
    \label{fig:ce-supcon}
\end{figure*}

\paragraph{Splits and evaluation protocol.}
Given the small number of recordings in the final subset, we use a leave-one-performance-out evaluation protocol. For each evaluation fold, one performance for each standard is held out as the test set, while remaining performances form the training set. Windows from the same held-out performance do not appear in the training set. When training supervised models, one additional random performance from the training side is reserved for validation and hyperparameter tuning. For the final reported fold results, hyperparameters were fixed and models were trained on the full training partition before being evaluated on the held-out performances.

\paragraph{Evaluation Metrics.}
We report three metrics based on classification and retrieval.
Window accuracy measures Top-1 classification of individual 10-second windows.
Performance Top-1 accuracy aggregates predictions over all windows from a held-out performance and evaluates the Top-1 standard prediction, corresponding to the strict standard identification setting.
Performance Top-5 reports whether the true standard appears among the five highest-scoring standards after performance-level aggregation, which better reflects a retrieval-based use case for jazz standard recognition.
We include window accuracy mainly as a diagnostic metric, since many individual windows may be noisy, or may not contain sufficient melodic or harmonic evidence for a given standard.

\section{Methods}

\paragraph{From-scratch Harmonic CNN.}
As a supervised baseline trained directly on the subset, we train a Harmonic CNN-style \cite{won2020harmonic} model on harmonic time-frequency representations of short audio windows. The model predicts a standard label for each window. At test time, we average window-level class probabilities over each held-out performance and choose the highest-scoring standard. 

\paragraph{Frozen pretrained embeddings.}
For this task, we extract frozen audio embeddings using MERT \cite{li2024mert} and MuQ \cite{zhu2025muq}. MERT-v1-95M\footnote{https://huggingface.co/m-a-p/MERT-v1-95M} is a pretrained musical understanding model with 95M parameters, while MuQ\footnote{https://huggingface.co/OpenMuQ/MuQ-large-msd-iter} is a similar larger model with 300M parameters with an open-weight version trained on the Million Song Dataset \cite{bertinmahieux2011million}.
For each window, hidden states are mean-pooled over time and concatenated across selected layers.
We then train lightweight classifiers on top of these frozen embeddings. Linear probing tests whether standard identity is linearly separable in the representation, while a small MLP classification head is used to test whether a nonlinear classifier improves the performance of the probe.

\paragraph{kNN-based retrieval.}
Given that performances from the same standard should follow similar melodies and underlying harmonies in theory, we also evaluate nearest-neighbor retrieval. Here, we build a reference set from embeddings from 10-second windows from the training split with MERT and MuQ encoders and L2-normalize each embedding. For each query window, we retrieve the top-$k$ nearest reference windows and aggregate their standard labels using temperature-scaled weighted voting with cosine similarity. We obtain performance-level scores by averaging window-level scores across all query windows.

\section{Results}
\begin{table*}[t]
\caption{Standard recognition results on the 16-standard JTD subset with 10-second windows. Random choice yields Top-1 accuracy of $1/16=0.0625$ and Top-5 accuracy of $5/16=0.3125$. We report mean $\pm$ standard deviation across folds.}
\label{tab:main_results}
\centering
\begin{tabular}{llccc}
\toprule
Method & Representation & Window Acc. & Perf. Top-1 & Perf. Top-5 \\
\midrule
Harmonic CNN \cite{won2020harmonic} & spectrogram & 0.034 $\pm$ \small{0.012} & 0.031 $\pm$ \small{0.036} & 0.359 $\pm$ \small{0.079} \\
\midrule
MERT-v1 \cite{li2024mert} & & & & \\
w/ linear probe & frozen embedding & 0.074 $\pm$ \small{0.056} & \textbf{0.094} $\pm$ \small{0.081} & 0.359 $\pm$ \small{0.139} \\
w/ MLP probe & frozen embedding & 0.096 $\pm$ \small{0.078} & \textbf{0.094} $\pm$ \small{0.091} & 0.422 $\pm$ \small{0.164} \\
w/ kNN retrieval ($k=5$) & frozen embedding & 0.066 $\pm$ \small{0.065} & 0.063 $\pm$ \small{0.051} & 0.359 $\pm$ \small{0.180} \\
\midrule
MuQ \cite{zhu2025muq} & & & & \\
w/ linear probe & frozen embedding & 0.085 $\pm$ \small{0.030} & 0.078 $\pm$ \small{0.031} & \textbf{0.469} $\pm$ \small{0.149} \\
w/ MLP probe & frozen embedding & \textbf{0.108} $\pm$ \small{0.068} & 0.078 $\pm$ \small{0.060} & 0.438 $\pm$ \small{0.102} \\
w/ kNN retrieval ($k=5$) & frozen embedding & 0.060 $\pm$ \small{0.058} & 0.078 $\pm$ \small{0.079} & 0.359 $\pm$ \small{0.107} \\
\bottomrule
\end{tabular}
\end{table*}

We compare the results of model training on the filtered dataset in Table~\ref{tab:main_results}. In particular, from the results we observe the following interesting properties:
\paragraph{From-scratch training overfits on the curated subset.}
The Harmonic CNN \cite{won2020harmonic} baseline trains well but its held-out performance remains close to chance-level under both Top-1 and Top-5 accuracy metrics, with worse-than-random performance on Top-1 accuracy. This suggests that the model learns performance-specific or recording-specific structure rather than a standard-invariant representation. This behavior is expected given the small number of independent performances in the dataset and the relative difficulty of the standard identification problem. The HCNN also shows low window-level accuracy across the test set, which supports these observations.

\paragraph{Pretrained embeddings consistently improve Top-$k$ recognition.}
Linear and MLP probes on MERT and MuQ embeddings improve over the from-scratch HCNN baseline and over random prediction odds, especially in the performance-level Top-5 accuracy metric. This indicates that audio foundation models indeed learn useful general-purpose representations for improved standard recognition, even when Top-1 recognition remains difficult. Although we assumed that the relative high dimensionality of the embeddings may be noisy, in our preliminary experiments dimensionality reduction with PCA before probing did not consistently improve these results. 

\paragraph{Retrieval is promising but naive retrieval is limited by performer identity.}
Nearest-neighbor retrieval eliminates the need for fine-tuning on the dataset toward a Shazam-like system because new standards can be added by inserting embeddings into the reference database. However, retrieval errors show a particular tendency: nearest neighbors are often drawn from the same trio or performer but correspond to different standards. This suggests that the embedding space of pretrained models preserves performer or recording similarity in addition to jazz standard identity. We revisit this idea in Section~\ref{sec:ablation}.

\begin{table*}[t]
\caption{Standard recognition results with varying window length. We report mean $\pm$ standard deviation across folds.}
\label{tab:20s_results}
\centering
\begin{tabular}{llccc}
\toprule
Method & Window Len. & Window Acc. & Perf. Top-1 & Perf. Top-5 \\
\midrule
MERT-v1 \cite{li2024mert} + linear probe & 10s & 0.074 $\pm$ \small{0.056} & 0.094 $\pm$ \small{0.081} & 0.359 $\pm$ \small{0.139} \\
 & 20s  & 0.065 $\pm$ \small{0.054} & 0.078 $\pm$ \small{0.060} & 0.375 $\pm$ \small{0.161} \\
MuQ \cite{zhu2025muq} + linear probe & 10s & 0.085 $\pm$ \small{0.030} & 0.078 $\pm$ \small{0.031} & \textbf{0.469} $\pm$ \small{0.149} \\
 & 20s  & 0.061 $\pm$ \small{0.015} & 0.063 $\pm$ \small{0.016} & 0.453 $\pm$ \small{0.107} \\
\midrule
MERT-v1 \cite{li2024mert} + MLP probe & 10s & 0.096 $\pm$ \small{0.078} & 0.094 $\pm$ \small{0.091} & 0.422 $\pm$ \small{0.164} \\
 & 20s  & 0.088 $\pm$ \small{0.065} & 0.094 $\pm$ \small{0.081} & 0.453 $\pm$ \small{0.164} \\
MuQ \cite{zhu2025muq} + MLP probe  & 10s &0.108 $\pm$ \small{0.068} & 0.078 $\pm$ \small{0.060} & 0.438 $\pm$ \small{0.102} \\
& 20s & \textbf{0.113} $\pm$ \small{0.090} & \textbf{0.125} $\pm$ \small{0.088} & \textbf{0.469} $\pm$ \small{0.063} \\
\bottomrule
\end{tabular}
\end{table*}

\begin{table*}
\caption{Retrieval-based standard recognition results with 10-second windows. }
\label{tab:supcon_results}
\centering
\begin{tabular}{llccc}
\toprule
Method & Same Group Freq. & Window Acc. & Perf. Top-1 & Perf. Top-5 \\
\midrule
MERT-v1 \cite{li2024mert} & & & & \\
w/ kNN probe ($k=5$) & 0.336 & 0.066 $\pm$ \small{0.065} & \textbf{0.063} $\pm$ \small{0.051} & 0.359 $\pm$ \small{0.180} \\
w/ kNN + SupCon ($\lambda=0.2$) & \textbf{0.109} &\textbf{ 0.081} $\pm$ \small{0.078} & \textbf{0.063} $\pm$ \small{0.051} & \textbf{0.469} $\pm$ \small{0.120} \\
\midrule
MuQ \cite{zhu2025muq} & & & & \\
w/ kNN probe ($k=5$) & 0.328 & 0.060 $\pm$ \small{0.058} & 0.078 $\pm$ \small{0.079} & 0.359 $\pm$ \small{0.107} \\
w/ kNN + SupCon ($\lambda=0.2$) & \textbf{0.156} & \textbf{0.095} $\pm$ \small{0.066} & \textbf{0.109} $\pm$ \small{0.060} & \textbf{0.438} $\pm$ \small{0.072} \\

\bottomrule
\end{tabular}
\end{table*}

\section{Ablations}
\label{sec:ablation}

\paragraph{Effect of window length.} With the relatively long duration of solos in performances from the dataset and low window accuracies from methods in Table~\ref{tab:main_results}, we investigate whether using a longer window length helps overall recognition for pretrained models. Albeit minor and model-specific, results from Table~\ref{tab:20s_results} show that increasing the window length most notably increases the Top-1 accuracy of the MuQ + MLP probe approach to a value of $0.125$, while being otherwise uninformative for other configurations.

\paragraph{Retrieval failures and improvements.} On investigation of the window-based retrieval approach from pretrained embeddings, we notice a particular trend: during retrieval, query windows from a given performance are often matched to windows from the same trio or performer but a different standard. For example, query sections of \emph{Autumn Leaves} from Ahmed Jamal might retrieve embeddings from Ahmed Jamal's \emph{Someday My Prince Will Come}, due to similarity in playing style, volume, or recording composition. Although these properties do represent a degree of similarity, this is not our target standard-focused similarity measure.

As a lightweight adaptation towards this problem, we experiment with the following pipeline. We keep the pretrained audio encoder frozen and train only a lightweight projection model on top of its window-level embeddings. Given a 10-second audio window, we first extract a fixed embedding using MERT or MuQ. This embedding is passed through a two-layer projection MLP to obtain a normalized representation $z_i$, and an MLP
 classification head maps $z_i$ to logits over the set of jazz standards. The model is trained with a combined objective:
\begin{equation}
\mathcal{L} = \mathcal{L}_{\mathrm{CE}} + \lambda \mathcal{L}_{\mathrm{SupCon}},
\end{equation}
where $\mathcal{L}_{\mathrm{CE}}$ denotes standard cross-entropy between the MLP head and the target standard label and $\mathcal{L}_{\mathrm{SupCon}}$ represents a supervised contrastive objective \cite{khosla2020supcon}. For a mini-batch of projected embeddings $\{z_i\}_{i=1}^N$, the supervised contrastive loss for anchor $i$ is
\begin{equation}
\begin{aligned}
\mathcal{L}_{\mathrm{SupCon},i}
&=
-\frac{1}{|P(i)|}
\sum_{p \in P(i)}
\log
\frac{\exp(s_{ip}/\tau)}
     {\sum_{a \in A(i)} \exp(s_{ia}/\tau)}, \\
s_{ij}
&= \operatorname{sim}(\mathbf{z}_i,\mathbf{z}_j),
\end{aligned}
\label{eq:supcon}
\end{equation}

  where $P(i)$ is the set of positive examples for anchor $i$, $A(i)=\{1,\dots,N\}\setminus{i}$, $\mathrm{sim}(\cdot,
  \cdot)$ is cosine similarity, and $\tau$ is a temperature parameter. In this setup, positives are windows with the
  same standard label but from a different performance, so that the projection is encouraged to learn cross-performance standard identity rather than recording similarity. After training, we discard the MLP head and use the learned projected embeddings for retrieval. Figure~\ref{fig:ce-supcon} shows the proposed pipeline.

We track the effectiveness of this method through the ``Same Group Frequency'' feature: while this ratio is high for vanilla kNN-based matching, we observe that this value drops significantly for kNN-based retrieval trained with the auxiliary supervised contrastive loss, and the overall Top-1 and Top-5 accuracy of the models appear to improve as well, as highlighted in Table~\ref{tab:supcon_results}. 
We set $\lambda=0.2$ based on preliminary validation experiments and keep it fixed across all folds to avoid any fold-specific tuning.

\section{Discussion and Future Work}

Our experiments suggest that jazz standard recognition is a challenging stress test for music representations. The difficulty is not merely the number of classes. In our curated dataset, many recordings appear to omit the head or begin in sections where the standard melody is not clearly present, which makes the task much harder than melody recognition. Consequently, a randomly sampled 10-second window may be considered difficult even for a human listener to identify. This makes window-level labels noisy: a solo passage labeled as \emph{Stella by Starlight} may contain little direct evidence of that standard, and we cannot accurately predict what is a ``good'' target accuracy for a trained model.

\paragraph{Future work.} The gap between Top-1 and Top-5 performance suggests that pretrained models often retrieve or classify within a plausible neighborhood but do not reliably rank the correct standard in first place. This motivates two future directions for this area of research: First, parameter efficient fine-tuning of musical foundation models instead of working with frozen embeddings is a very promising candidate for improving recognition. Second, further explicit analysis of the underlying melody and harmony with a given audio sample may further provide useful features towards better generalization of such recognition models.

\paragraph{Limitations.} Our methodology and evaluation use local audio windows rather than a full symbolic or harmonic representation of a performance. As a result it only has a limited access to long-range musical context, which is important as many standards share similar local windows but possess a characteristic global flow that is important for identification.
Conversely, many standards that share similar progressions (e.g., a 12-bar blues) are identifiable mainly through small but distinctive excerpts which can be rare relative to the duration of an entire performance. In this context, models might benefit from longer-context modeling, melody extraction and harmony identification for better results.
Lastly, our curated dataset is intentionally small and controlled, so the results should be interpreted as an exploratory benchmark rather than a definitive ranking of music foundation models.

\section{Conclusion}

In this study, we present an exploratory study of jazz standard recognition across performances using a curated subset of the Jazz Trio Database. Our results indicate that a from-scratch HCNN network overfits to training performances, while probing based on embeddings from audio encoding models such as MERT and MuQ provide stronger results due to their pretraining. Our retrieval analysis suggests that the group identity influences embedding-based similarity and retrieval. Ultimately, our results indicate that jazz standard recognition is a realistic but difficult setting for evaluating whether music representations encode tune identity rather than only acoustic or performance similarity.

\bibliography{main}
\bibliographystyle{icml2026}

\end{document}